\documentclass[letterpaper,twocolumn,english,prl]{revtex4}

\usepackage[T1]{fontenc}
\usepackage[utf8x]{inputenc}
\usepackage{color}
\usepackage{amsmath}
\usepackage{stackrel}
\usepackage{graphicx}
\usepackage{esint}

\makeatletter

\@ifundefined{textcolor}{}
{%
 \definecolor{BLACK}{gray}{0}
 \definecolor{WHITE}{gray}{1}
 \definecolor{RED}{rgb}{1,0,0}
 \definecolor{GREEN}{rgb}{0,1,0}
 \definecolor{BLUE}{rgb}{0,0,1}
 \definecolor{CYAN}{cmyk}{1,0,0,0}
 \definecolor{MAGENTA}{cmyk}{0,1,0,0}
 \definecolor{YELLOW}{cmyk}{0,0,1,0}
}

\@ifundefined{definecolor}
 {\usepackage{color}}{}

\makeatother

\usepackage{babel}
\begin{document}

\title{The Michaelis-Menten reaction scheme~\\
{\footnotesize{}\medskip{}
}as a unified approach towards the optimal restart problem }

\author{{\normalsize{}Tal }\textcolor{black}{\normalsize{}Rot}{\normalsize{}bart$^{\dagger,*}$,
Shlomi Reuveni$^{\ddagger,*}$, and Michael Urbakh$^{\dagger}$}}

\affiliation{\noindent \textit{$^{\dagger}$School of Chemistry, Tel-Aviv University,
Tel-Aviv 69978, Israel}}

\affiliation{\textit{$^{\ddagger}$Department of Systems Biology, Harvard Medical
School, 200 Longwood Avenue, Boston, Massachusetts 02115, USA.}}

\affiliation{$^{*}$\textit{T. Rotbart and S. Reuveni had equal contribution to
this work.}}
\begin{abstract}
We study the effect of restart, and retry, on the mean completion
time of a generic process. The need to do so arises in various branches
of the sciences and we show that it can naturally be addressed by
taking advantage of the classical reaction scheme of Michaelis \&
Menten. Stopping a process in its midst—only to start it all over
again—may prolong, leave unchanged, or even shorten the time taken
for its completion. Here we are interested in the optimal restart
problem, i.e., in finding a restart rate which brings the mean completion
time of a process to a minimum. We derive the governing equation for
this problem and show that it is exactly solvable in cases of particular
interest. We then continue to discover regimes at which solutions
to the problem take on universal, details independent, forms which
further give rise to optimal scaling laws. The formalism we develop,
and the results obtained, can be utilized when optimizing stochastic
search processes and randomized computer algorithms. An immediate
connection with kinetic proofreading is also noted and discussed.
\end{abstract}
\maketitle
When engaged in a specific task for a time period that extends beyond
our initial expectations, we are constantly faced with two alternatives---either
keep on going or stop everything and start anew. Every now and then
we opt for the latter, hoping that a fresh start will break-off an
unproductive course of action and expedite the completion of the task
at hand. This decision could, however, turn out to be counter-productive---nipping
an awaited, but unforeseen, finale in the bud. To restart, or not
to restart, that is therefore the question. 
\begin{figure}[t]
\begin{centering}
\includegraphics[scale=0.43]{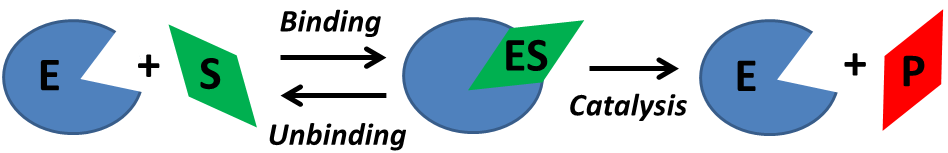}
\par\end{centering}

\protect\caption{\label{Fig1}\textbf{\textcolor{black}{Color Online.}}\textbf{\small{}
The Michaelis-Menten reaction scheme.} An enzyme $E$ can reversibly
bind a substrate $S$ to form a complex $ES$. The substrate can then
be converted by the enzyme to form a product $P$ or, alternatively,
unbind. The conversion of the substrate to a product is manifested
via the process of enzymatic catalysis. Following either catalysis,
or unbinding, the enzyme is free to act on additional substrate molecules.}
\end{figure}

Not at all unique to our everyday lives, a ``dilemma'' similar to
the one described above is relevant to virtually any physical, chemical,
or biological process that can be restarted. Most notably, restart
(or unbinding) is an integral part of the renown Michaelis-Menten
Reaction Scheme (MMRS) illustrated in Fig. 1 \cite{MM}. Originally
devised to describe enzymatic catalysis, the MMRS has attracted on-growing
scientific interest for more than a century \cite{MMSI}. Indeed,
nature is full with an astonishing variety of Michaelian processes.
DNA-DNA hybridization, antigen-antibody binding, and various other
molecular processes can all be described by the MMRS \cite{Lehninger}.
That and more, the simplicity and generality of the scheme have rendered
it widely applicable and it is now used to describe anything from
heterogeneous catalysis \cite{Hetro1,Hetro2,Hetro3} to in vivo target
search kinetics \cite{In Vivo Search1}. As a matter of fact, one
can easily convince himself that\emph{ any }first passage time (FPT)
process \cite{Redner}---be it the time to target of a simple Brownian
particle, or that of more sophisticated stochastic processes \cite{Non-Brownian1,Non-Brownian2,FPT in complex invariant media,Geometry-controlled kinetics}
and random searchers \cite{Search1,Search2,Search3}---can become
subject to restart \cite{Restart1,Restart2,Restart3,Restart4,Restart5,Restart6,Restart7}
and is then naturally accommodated by the MMRS. Wishing to acquire
a unified view on restart phenomena we identify the MMRS as an ideal
object of study. 

Central to our understanding of the MMRS is the Michaelis-Menten (MM)
equation \cite{MM}. This equation provides a hundred-year-old prediction
by which any increase in the rate of unbinding (restart) will inevitably
slow down the rate of enzymatic turnover or, equivalently, prolong
the completion time of any process that falls into the MMRS category.
Surprisingly, this prediction was never tested experimentally, but
rapid advancements in single-molecule techniques \cite{Ever fluctuating,Fernandez1,Triggering enzymatic activity with force}
have recently motivated us to question it from a theoretical perspective
\cite{The Role}. Contrary to the classical result, we have found
that unbinding may also facilitate the successful completion of a
reaction. In the emerging picture, a non-trivial solution to the restart
dilemma is given by an optimal unbinding rate which strikes the right
balance between the need to abort prolonged reaction cycles and the
need to avoid premature termination of ongoing ones. Similar observations
were made in the context of search processes \cite{Restart2} and,
in particular, in the context of DNA search \cite{Restart1} where
the authors analyzed facilitated diffusion \cite{FD} from a very
general perspective.\textcolor{red}{{} }The optimal restart rate depends,
however, on the full distribution of the underlying FPT process (catalysis
in the case of enzymatic reactions) \cite{The Role}, and the question
of what can be said about it in the general case remained open.

In this letter, we address the optimal restart problem within the
framework of the MMRS. First, the governing equation for this problem,
Eq. (\ref{3}) below, is derived and solved exactly in several cases
that are of particular interest. We then show that, in the general
case, there are two regimes at which solutions to the problem are
universal. These solutions are given in Eqs. (\ref{4}) and (\ref{5}).
The applicability of our approach is widespread as it allows one to
incorporate restart into an existing, generic, FPT problem in an almost
plug \& play manner. 

\textbf{The optimal restart problem. }In formulating the optimal restart
problem we adopt the terminology of enzymatic reactions (Fig. 1) and
consider a scenario in which the processes of binding, unbinding and
catalysis are all stochastic \cite{Def}. This probabilistic view
point, whose origins can be traced back to the work of Ninio \cite{Ninio},
has found one of its prime applications in the analysis of single
molecule experiments \cite{Single-molecule enzymic dynamics,The =00FB02uctuating enzyme,Flomenbom,Ever fluctuating},
and is now well established theoretically \cite{Kou,Cao1,Cao4}. Defining
the turnover rate $k_{turn}$ as the reciprocal of the turnover time---the
mean FPT required to complete the reaction cycle---we will be interested
in the turnover--unbinding interplay. 

When binding, unbinding, and catalysis times are exponentially distributed
with rates $k_{on}[S]$ ($[S]$ being the concentration of the substrate),
$k_{off}$, and $k_{cat}$ respectively---the single molecule MM equation
is attained \cite{Kou} 
\begin{equation}
k_{turn}=\frac{k_{cat}[S]}{[S]+K_{M}}\,,\label{1}
\end{equation}
with $K_{M}=(k_{off}+k_{cat})/k_{on}$ (note that, in contrast to
other rates in the MM equation, the turnover “rate” is not a rate
in the “exponential sense”). In this case, the memory-less property
of the exponential distribution asserts that the time remaining till
the completion of an ongoing catalytic step, given its age, is exponential
and statistically identical to that of a newly started catalytic step.
It is therefore clear (see $K_{M}$ above) that $k_{turn}$ is a monotonically
decreasing function of the rate $k_{off}$. 

Non-exponential time distributions are, however, quite common in a
variety of complex systems \cite{First Steps in Random Walks,non-exp1,non-exp2,non-exp3,non-exp4},
and it has recently been recognized that enzymes are no exception
in that regard \cite{Flomenbom,Ever fluctuating,Randomness par1,Randomness par2}.
This result is perhaps not surprising as catalysis is intrinsically
coupled to the enzyme's internal degrees of freedom via a complex
energy landscape \cite{Memory landscapes} which can give rise to
strong deviations from exponentiality and other anomalies \cite{Granek,Reuveni1,Reuveni2,Reuveni3,Reuveni4}.
Renewal theory can then be invoked to provide a generalized mathematical
treatment of the MMRS \cite{The Role}. A completely general analysis
of the turnover-unbinding interplay is then very hard, but progress
can be made if one narrows down to the case of exponentially distributed
unbinding times \cite{Kou,The Role}. Letting $k_{off}$ denote the
unbinding rate (assumed to be independent of the catalytic process),
and $f_{cat}(t)$ the probability density function (PDF) of a generally
distributed catalysis time $T_{cat}$, it is possible to show that
\cite{Kou,The Role} 
\begin{equation}
k_{turn}=\frac{\hat{f}_{cat}(k_{off})}{\left\langle T_{on}\right\rangle +k_{off}^{-1}\left[1-\hat{f}_{cat}(k_{off})\right]}\,,\label{2}
\end{equation}
where $\hat{f}_{cat}(k)=\overset{\infty}{\underset{0}{\int}}f_{cat}(t)e^{-kt}dt$
is the Laplace transform of $f_{cat}(t)$, and $\left\langle T_{on}\right\rangle $
is the mean of a generally distributed binding time. Equation (\ref{2})
extends the classical result of Michaelis \& Mente\textcolor{black}{n
and brings new physics}. Indeed, an interesting corollary of Eq. (\ref{2})
is the possibility of restart-facilitated-turnover, i.e., a regime
in which \emph{unproductive} unbinding events lead to accelerated
turnover \cite{The Role}. This type of counter-intuitive behavior
is categorically precluded by the classical MM equation and is therefore
considered ``non-classical'' or ``anomalous''. 

\begin{figure*}[t]
\includegraphics[scale=0.4]{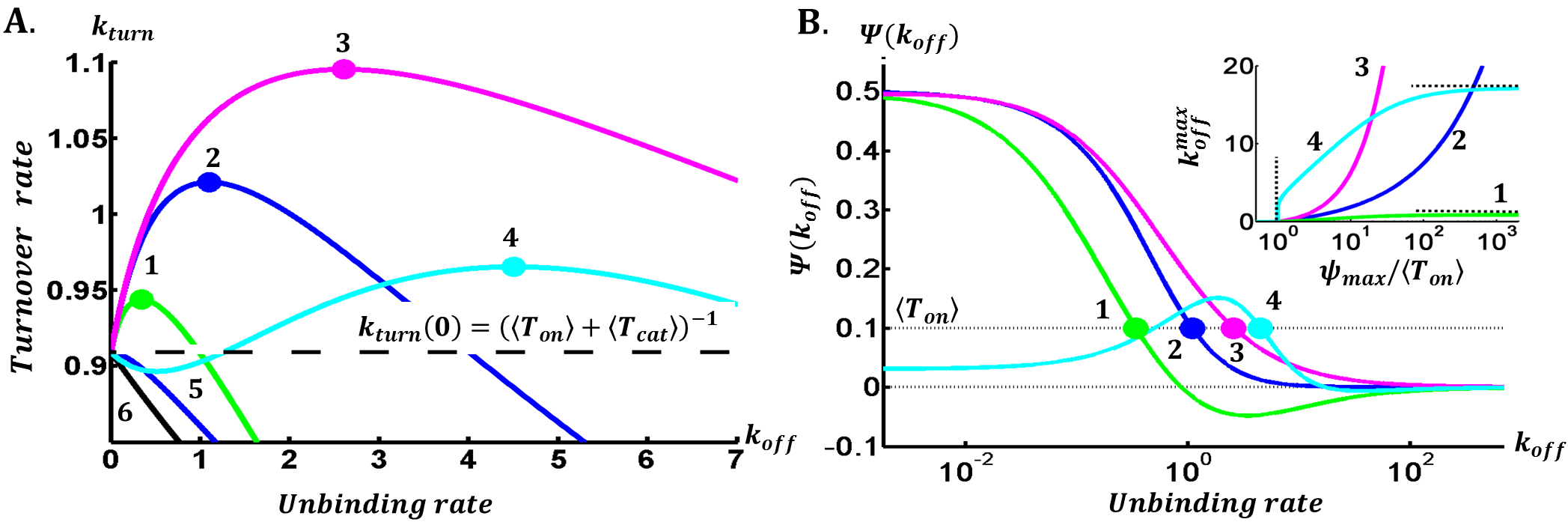}\protect\caption{\label{Fig2}\textbf{\textcolor{black}{Color Online.}}\textbf{ A.}
\textbf{Turn-over rate vs. unbinding (restart) rate for different
catalysis time distributions \cite{Supplemental Material}.} Contrary
to classical theory, non-monotonic dependencies are possible and the
classical upper limit on turnover rates (dashed) can be broken. The
distributions are numbered by: 1-Log normal, 2-Double exponential,
3-Weibul, 4-Double Erlang, 5-Double exponential, 6-Exponential. Heterogeneity
is observed despite the fact that $\left\langle T_{on}\right\rangle =0.1$
in all instances, all underlying time distributions share a mean of
$\left\langle T_{cat}\right\rangle =1$, and some even share the same
variance (\textcolor{black}{$\sigma^{2}(T_{cat})=2$ for }distributions
No. 1-3\textcolor{black}{)}. Maxima in turnover rates are denoted
by full circles. \textbf{B. A graphical illustration of the fundamental
equation of optimal restart (Eq. \ref{3}).} Extrema of the turnover
rate are obtained at points where $\Psi(k_{off})$ intersects the
mean binding time $\left\langle T_{on}\right\rangle $. Maxima from
panel (A) are once again denoted by full circles. Lines for distributions
No. 5-6, where $\Psi(k)\leq0$, are not drawn. Note that line No.
4 intersects $\left\langle T_{on}\right\rangle $ twice, first with
a positive slope (minimum of $k_{turn}$) and then with a negative
slope (maximum of $k_{turn}$). \textbf{Inset B.} \textbf{$k_{off}^{max}$
vs. $\Psi_{max}/\left\langle T_{on}\right\rangle $.} When $\left\langle T_{on}\right\rangle >\Psi_{max}$,
$k_{off}^{max}=0$. As $\left\langle T_{on}\right\rangle $ drops
$\Psi_{max}$, a maxima may develop gradually (lines No. 1-3) or abruptly
(line No. 4). As $\left\langle T_{on}\right\rangle $ approaches zero,
$k_{max}$ can either diverge (lines No. 2 and 3) or asymptotically
converge to a plateau (lines No. 1 and 4). }
\end{figure*}

In Fig. \ref{Fig2}A we use Eq. (\ref{2}) to plot $k_{turn}$ as
a function of $k_{off}$ for different catalysis time distributions
(CTDs). An asymptotic decay of $k_{turn}$ to zero at large $k_{off}$
directly follows from Eq. (\ref{2}), is common to all plots, and
is therefore not shown. At intermediate $k_{off}$, however, unbinding
can be either \emph{inhibitory} ($\partial k_{turn}/\partial k_{off}<0)$
or \emph{excitatory} ($\partial k_{turn}/\partial k_{off}>0)$ and
the surprise comes from the fact that the latter implies the\emph{
breaking} of the classical limit for maximal turnover rates: $k_{turn}(0)=\left(\left\langle T_{on}\right\rangle +\left\langle T_{cat}\right\rangle \right)^{-1}$.
Maximal turnover rates, and the unbinding rates at which they are
attained, can however vary considerably. What therefore determines
if unbinding will enhance turnover, and what sets the maximizing unbinding
rate $k_{off}^{max}$?

\textbf{The fundamental equation of optimal restart.} In order to
address the questions presented above we first derive a governing
equation for the optimal restart problem. Namely, we show \cite{Supplemental Material}
that the stationary points of the turnover rate are the solutions
of 
\begin{equation}
\Psi(k)\equiv\frac{\hat{f}_{cat}(k)(\hat{f}_{cat}(k)-1)}{k^{2}d\hat{f}_{cat}(k)/dk}-\frac{1}{k}=\left\langle T_{on}\right\rangle \,,\label{3}
\end{equation}
where the function $\Psi(k)$ is uniquely determined by the CTD. Moreover,
we show that $\Psi(k)$ has the following property: $\partial k_{turn}/\partial k_{off}>0\Longleftrightarrow\Psi(k)>\left\langle T_{on}\right\rangle $
and vice versa. Consequently, a local maximum of the turnover rate
is attained at an unbinding rate, $k_{off}^{max}$, which satisfies
$\Psi(k_{off}^{max})=\left\langle T_{on}\right\rangle $ and $\Psi'(k_{off}^{max})<0$. 

Equation (\ref{3}) clarifies the role of binding in our problem.
Restart and the initiation of a new turnover attempt, inevitably involve
a ``penalty''---the necessity to go through the binding process
all over again. Taking the perspective of turnover rate maximization,
lengthy binding times are hence a deterrent against restart and rapid
binding an incentive to it. In perfect accord with this intuition,
we note that a maximum in $k_{turn}$ will develop if and only if
$\left\langle T_{on}\right\rangle $ drops below the critical value
of $\Psi_{max}\equiv\underset{k>0}{max}\{\Psi(k)\}$. Indeed, since
$k_{turn}$ is positive and asymptotically decays to zero as $k_{off}\rightarrow\infty$,
$\Psi(k)$ will eventually intersect any level in the range $\Psi_{max}>\left\langle T_{on}\right\rangle >0$
with a negative slope (provided $\Psi_{max}>0$) \cite{Converse}.
In particular, a maximum in $k_{turn}$ will develop whenever $\left\langle T_{on}\right\rangle $
drops below $\Psi_{0}\equiv\Psi(0)$---an observation that will come
in handy later on. These properties of $\Psi(k)$ are graphically
illustrated in Fig. \ref{Fig2}B. 

\textbf{A classical example and a family of exactly solvable cases.
}The importance of Eq. (\ref{3}) cannot be overstated as it allows
one to find optimal restart rates for generic FPT processes. In order
to demonstrate the power of this formalism, we will now reanalyze
a problem studied by Evans \& Majumdar in \cite{Restart2}. Consider
a particle searching for a stationary target via one dimensional diffusion.
Setting the initial distance between the particle and target to $L$
and the diffusion coefficient to $D$, it has long been known that
the mean FPT to the target diverges. What happens, however, if the
particle is returned (restarted) to its initial position with some
given rate $k_{off}$ (assume $\left\langle T_{on}\right\rangle =0$)?
Well, since the FPT distribution of the original problem is known
to be given by $\hat{f}_{cat}(k)=e^{-\sqrt{kL^{2}/D}}$ (Laplace space
representation of the Lévy-Smirnov distribution), it immediately follows
that the mean FPT of the restarted problem is given by $k_{turn}^{-1}$
in Eq. (\ref{2})---and we further note that it is finite for any
positive restart rate! In fact, this is true for any $\hat{f}_{cat}(k)$,
and as long as $\left\langle T_{on}\right\rangle $ is finite, but
is particularly striking when the underlying FPT process is equipped
with an infinite mean.\textcolor{red}{{} }Moreover, by solving Eq. (\ref{3})
one can see that $k_{off}^{max}=(z^{*})^{2}D/L^{2}$, where $z^{*}\simeq1.59362...$
is the solution to $z/2=1-e^{-z}$. Clearly, the same modus operandi
can also be used to study the effect of restart on many other FPT
classics \cite{Redner}. In particular, one could readily generalize
the above example for the one sided Lévy distribution $\hat{f}_{cat}(k)=e^{-\left(\tau k\right){}^{\alpha}}$
$(0<\alpha<1)$ to obtain $k_{off}^{max}=(z^{*})^{1/\alpha}/\tau$,
where $z^{*}$ is the solution to $\alpha z=1-e^{-z}$.

Analytical solutions to Eq. (\ref{3}) are hard to find. It is thus
interesting to note that whenever the Laplace transform of the CTD
has the following form: $\hat{f}_{cat}(k)=(1+ak)/(1+bk+ck^{2})$ (for
some constants $a$, $b,$ and $c$)---Eq. (\ref{3}) reduces to a
quadratic and is hence exactly solvable. A particular example in this
category is the Exponential distribution, $f_{cat}(t)=\lambda e^{-\lambda t}\,\,(\lambda>0)$,
for which $a=c=0$, $b=\frac{1}{\lambda}$, and $\Psi(k)=0$. As another
example, think of the Double-Exponential distribution, $f_{cat}(t)=p\lambda_{1}e^{-\lambda_{1}t}+(1-p)\lambda_{2}e^{-\lambda_{2}t}\,\,(0<p<1,\lambda_{1}>0,\lambda_{2}>0)$,
for which $a=\frac{1}{\lambda_{1}}+p\left(\frac{1}{\lambda_{2}}-\frac{1}{\lambda_{1}}\right)$,
$b=\frac{1}{\lambda_{2}}+\frac{1}{\lambda_{1}}$, $c=\frac{1}{\lambda_{1}\lambda_{2}}$,
and {\small{}$\Psi(k)=\left(\frac{p\lambda_{1}+(1-p)\lambda_{2}}{(1-p)p\left(\lambda_{1}-\lambda_{2}\right){}^{2}}k^{2}+\frac{2\left(\lambda_{1}\lambda_{2}\right)}{(1-p)p\left(\lambda_{1}-\lambda_{2}\right){}^{2}}k+\frac{\lambda_{1}\lambda_{2}\left((1-p)\lambda_{1}+p\lambda_{2}\right)}{(1-p)p\left(\lambda_{1}-\lambda_{2}\right){}^{2}}\right)^{-1}\,.$}
Finally, and perhaps most importantly, consider the class of distributions
which do not fall into the above-mentioned form, but can rather be
\emph{asymptotically} approximated by it. As we hereby show, the basin
of attraction for this class is wide---rendering asymptotic solutions
to the optimal restart problem (almost) universal. 

\textbf{Universal behavior at }$\left\langle T_{on}\right\rangle \approx\Psi_{0}$\textbf{.}
When $\left\langle T_{on}\right\rangle $ approaches $\Psi_{0}$,
``small $k$'' solutions to Eq. (\ref{3}) are anticipated provided
$\Psi_{0}>0$ (see Fig. \ref{Fig2}B). One can then try and approximate
$\Psi(k)$, at small $k$, considering that in this limit $\hat{f}_{cat}(k)\simeq1+\stackrel[n=1]{m}{\sum}\frac{M_{n}(-k)^{n}}{n!}$,
where $M_{n}\equiv\left\langle T_{cat}^{n}\right\rangle $ is the
$n-th$ moment of the CTD. It can then be verified, utilizing the
definition of $\Psi(k)$, that this expansion must be carried out
to third order in $k$ ($m=3$)---if it were to correctly capture
$\Psi(k)$ to first order. However, under direct substitution of such
an expansion into $\Psi(k)$, Eq. (\ref{3}) becomes a forth order
equation and further analytical advancement becomes extremely cumbersome. 

To circumvent this difficultly, we make use of the widely applied
Padé approximation scheme \cite{Pade} and try $\hat{f}_{cat}(k)\simeq(1+ak)/(1+bk+ck^{2})$.
Doing so, we note that this approximation: (i) can be made exact to
third order in $k$ by proper choice of the constants $a$, $b,$
and $c$ \cite{Supplemental Material}; (ii) decays to zero as $k\rightarrow\infty$---as
required from a Laplace transform, and in sharp contrast to the divergences
of any power series expansion; and (iii) renders the solution to Eq.
(\ref{3}) immediate as it gives for $k\ll1$ \cite{Supplemental Material}:
$\Psi(k)\simeq\frac{\Psi_{0}}{1+2R_{0}\left(\Psi_{0}k\right)+R_{0}(1+R_{0})\left(\Psi_{0}k\right)^{2}}\,,$
where $\Psi_{0}=(M_{2}-2M_{1}^{2})/2M_{1}$ and $R_{0}=\frac{2M_{1}M_{3}-3M_{2}^{2}}{3\left(M_{2}-2M_{1}^{2}\right)^{2}}$.
We now see that in this limit solutions to the optimal unbinding problem
are insensitive to fine details of the CTD as they are governed by
$\Psi_{0}$ and $R_{0}$ only. 

As we have previously observed, the \emph{introduction} of unbinding
is asserted to speed up turnover whenever $\left\langle T_{on}\right\rangle <\Psi_{0}$
since this implies $\frac{\partial k_{turn}}{\partial k_{off}}|_{k_{off}=0}>0$.
The newly derived expression for $\Psi_{0}$ allows us to interpret
this result probabilistically. Indeed, setting $k_{off}$ to zero,
it is easy to see that $\left\langle T_{on}\right\rangle <\Psi_{0}$,
if and only if, the mean duration, $\left\langle T_{on}\right\rangle +\left\langle T_{cat}\right\rangle $,
of a new turnover cycle drops below the mean \emph{residual} duration,
$\frac{1}{2}\left\langle T_{cat}^{2}\right\rangle /\left\langle T_{cat}\right\rangle $,
of an ongoing catalytic step \cite{Gallager}. Unbinding will then
have an excitatory effect but two distinct scenarios should nevertheless
be told apart. 

When $R_{0}>0$, $\Psi'(0)<0$ (e.g., lines No. 1-3 in Fig. \ref{Fig2}B)
and, as $\left\langle T_{on}\right\rangle $ approaches $\Psi_{0}$
from above and crosses over to its other side, a \emph{maximum} of
the turnover rate \emph{gradually develops} at 

{\small{}
\begin{equation}
k_{off}^{max}\simeq\frac{1}{\left(1+R_{0}\right)\Psi_{0}}\left(\sqrt{\left(1+\frac{1}{R_{0}}\right)\frac{\Psi_{0}}{\left\langle T_{on}\right\rangle }-\frac{1}{R_{0}}}-1\right)\,.\label{4}
\end{equation}
}In particular, setting $\triangle=\left\langle T_{on}\right\rangle ^{-1}-\Psi_{0}^{-1}$,
we observe that to first order $k_{off}^{max}\simeq\frac{\triangle}{2R_{0}}$.
This characteristic dependence is further discussed in Fig. S1 \cite{Supplemental Material}. 

On the other hand, when $R_{0}<0$ (e.g. line No. 4 in Fig. \ref{Fig2}B),
$\Psi'(0)>0$, and $\Psi(k)$ has a local maxima at some $k^{*}>0$.
Then, as $\left\langle T_{on}\right\rangle $ first hits $\Psi(k^{*})$
from above, both a minimum and a maximum of $k_{turn}$ \emph{abruptly
appear} (see ``jump'' in $k_{off}^{max}$, inset of Fig. \ref{Fig2}B).
As $\left\langle T_{on}\right\rangle $ continues to decrease, these
two extrema drift apart and it is important to observe that the small
$k$ solution to Eq. (\ref{3}) is then a \emph{minimum}, rather than
a maximum. As $\left\langle T_{on}\right\rangle $ drops below $\Psi_{0}$,
this minimum necessarily disappears---leaving behind a maximum of
$k_{turn}$ at a point $k_{off}^{max}$ which is \emph{strictly} \emph{separated}
from zero. Before moving forward, we note in passing that $R_{0}<0$
if and only if the residual duration of an ongoing catalytic step
has a coefficient of variation that is smaller than unity. 

We end this section by noting that when the catalysis time distribution
is ``heavy tailed''---as happens in a wide variety of FPT problems---either
one of its first two moments can diverge. An abrupt phase transition
is then observed---$\Psi(k\rightarrow0)=\infty$ and the \emph{introduction}
of unbinding is asserted to speed up turnover regardless of $\left\langle T_{on}\right\rangle $.
The asymptotic behavior of $k_{off}^{max}$ at high values of $\left\langle T_{on}\right\rangle $
then depends on the tail of the catalysis time distribution ($t\rightarrow\infty$),
and it can be shown \cite{Supplemental Material} that for $f(t)\sim t^{-(1+\alpha)}$
with $0<\alpha<1$ ($1<\alpha<2$), $k_{off}^{max}\sim\left\langle T_{on}\right\rangle ^{-1}$
($k_{off}^{max}\sim\left\langle T_{on}\right\rangle ^{-1/(2-\alpha)}$).
One example for this type of behavior is the above-mentioned case
of diffusion mediated search for which $\alpha=1/2$. 
\begin{figure}
\includegraphics[scale=0.16]{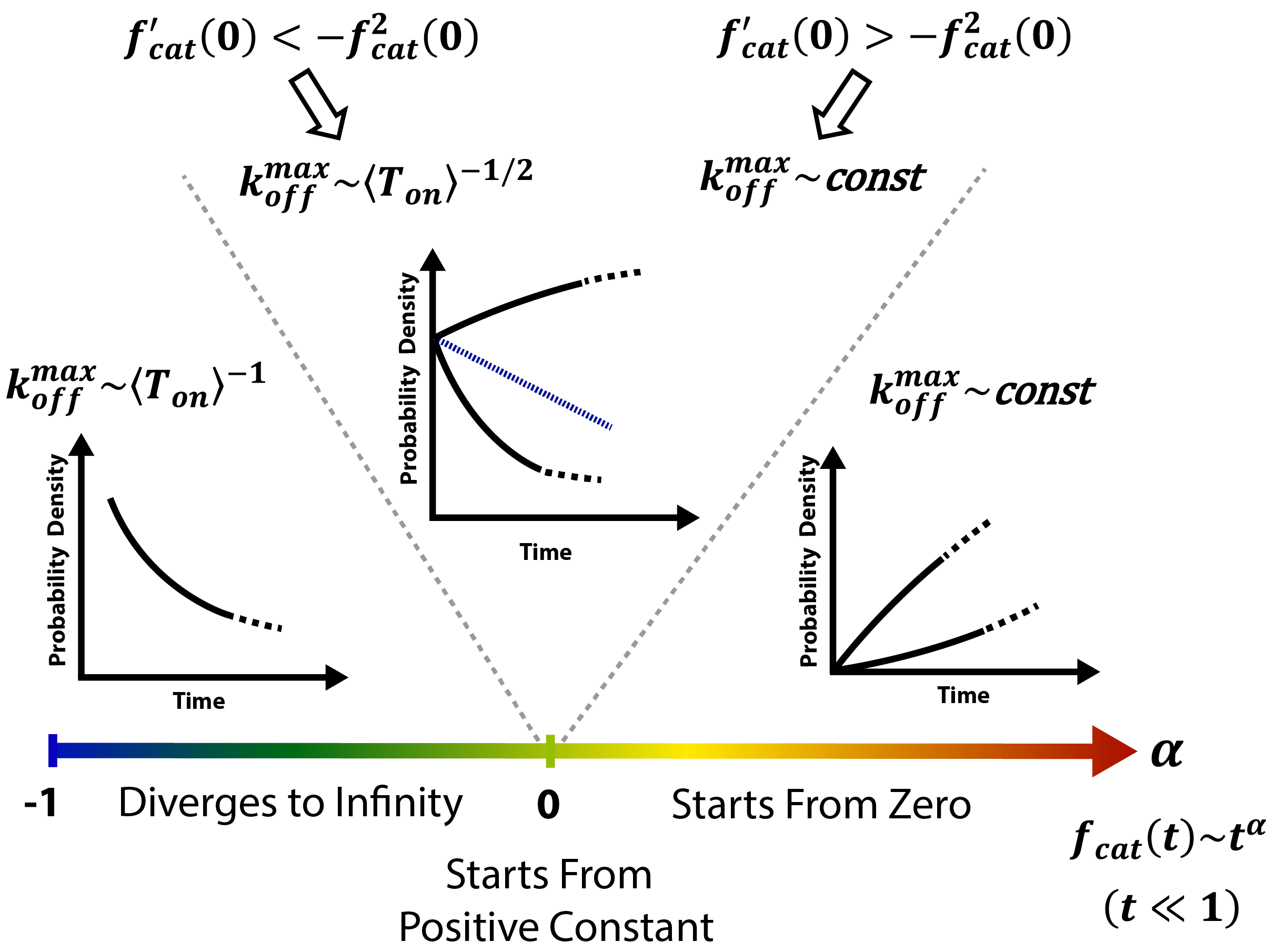}\protect\caption{\label{Fig3}\textbf{\textcolor{black}{Color Online.}}\textbf{ Asymptotics
of the optimal restart problem, at fast binding times, is governed
by the behavior of the catalysis time distribution near the origin.}}
\end{figure}

\textbf{Universal behavior at fast binding times.} When $\left\langle T_{on}\right\rangle $
approaches zero, ``large $k$'' solutions to Eq. (\ref{3}) are
anticipated provided $\Psi(k)$ is asymptotically positive (see Fig.
\ref{Fig2}B). The behavior of $\Psi(k)$ in this limit is governed
by the behavior of $f_{cat}\left(t\right)$ at short times and we
progress by assuming that $f_{cat}\left(t\right)\sim t^{\alpha}\,\,(t\ll1)$.
Three different regimes, illustrated in Fig. \ref{Fig3}, are then
noteworthy \cite{Supplemental Material}. When $-1<\alpha<0,$ $\Psi(k)$
approaches zero from above as $\sim k^{-1}$, and $k_{off}^{max}\sim\left\langle T_{on}\right\rangle ^{-1}$.
On the other hand, when $\alpha>0,$ $\Psi(k)$ approaches zero from
below as $\sim-k^{-1}$, there are no ``large $k$'' solutions to
Eq. (\ref{3}), and $k_{off}^{max}|_{\left\langle T_{on}\right\rangle =0}$
is finite (can be zero). 

The case $\alpha=0$ is a bit more delicate. Assuming $f_{cat}\left(t\right)$
has a Taylor expansion near the origin we denote $\omega_{n}=n!\left.\left(\left(-\frac{\partial}{\partial t}\right)^{n-1}f_{cat}\left(t\right)\right)\right|_{t=0}$
and note that by construction $\omega_{1}=f_{cat}\left(0\right)>0.$
Implementing a treatment similar to one given in the previous section
we find for $k\gg1$ \cite{Supplemental Material}: $\Psi(k)\simeq\frac{\chi_{\infty}/k^{2}}{1+2R_{\infty}\left(\chi_{\infty}/k\right)+R_{\infty}(1+R_{\infty})\left(\chi_{\infty}/k\right)^{2}}\,,$
where $R_{\infty}=\frac{2\omega_{1}\omega_{3}-3\omega_{2}^{2}}{3(\omega_{2}-2\omega_{1}^{2})^{2}}$
and $\chi_{\infty}=(\omega_{2}-2\omega_{1}^{2})/2\omega_{1}$ (compare
with $\Psi(k)$ for $k\ll1$ above). A large $k$ solution to Eq.
(2) is then found only when $f_{cat}(t)$ decreases steeply enough
near the origin, i.e., when $f_{cat}^{'}\left(0\right)<-f_{cat}^{2}\left(0\right)\Longleftrightarrow\chi_{\infty}>0$,
and is given by 

\begin{equation}
k_{off}^{max}=\chi_{\infty}\left(\sqrt{\frac{1}{\left\langle T_{on}\right\rangle \chi_{\infty}}-R_{\infty}}-R_{\infty}\right)\,.\label{5}
\end{equation}
In particular, note that in this case $k_{off}^{max}\sim\left\langle T_{on}\right\rangle ^{-1/2}$.
On the other hand, when $\chi_{\infty}<0$, we once again find that
$k_{off}^{max}|_{\left\langle T_{on}\right\rangle =0}$ is a constant. 

\begin{figure}
\includegraphics[scale=0.38]{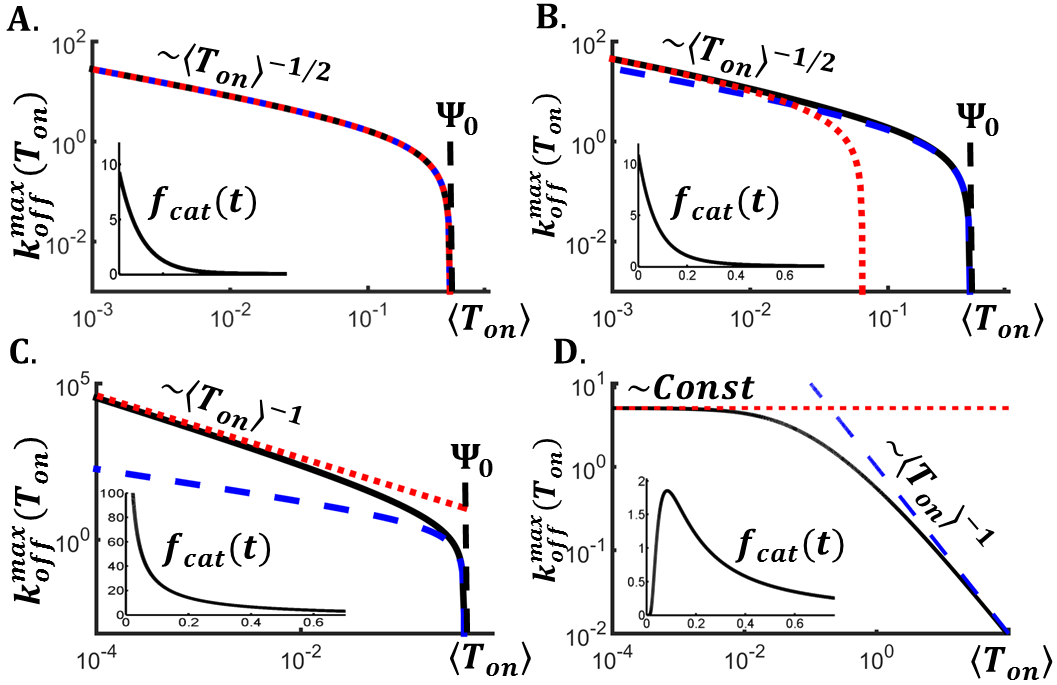}\protect\caption{\label{Fig4}\textbf{\textcolor{black}{Color Online.}}\textbf{ Optimal
unbinding (restart) rate vs. mean binding time for different catalysis
time distributions. }The black (full), blue (dashed), and red (dotted)
curves correspond to $k_{off}^{max}$, and to asymptotic approximations
of $k_{off}^{max}$ in the $\left\langle T_{on}\right\rangle \rightarrow\Psi_{0}$,
and $\left\langle T_{on}\right\rangle \rightarrow0$, limits respectively.
Distributions (A-Double Exponential, B-Triple Exponential, C-Gamma,
D-Lévy-Smirnov) are drawn at the bottom left corner of each panel
\textbf{\cite{Supplemental Material}}. Dashed blue lines in panels
(A-C) are drawn according to Eq. (\ref{4}). A phase transition occurs
in panel (D) where $\left\langle T_{cat}\right\rangle =\infty$ and
asymptotics is governed by a $\sim\left\langle T_{on}\right\rangle ^{-1}$
scaling law. Dotted red lines in panels A \& B are drawn according
to Eq. (\ref{5}). A different behavior is found in panels C \& D---in
accord with the $t\rightarrow0$ asymptotics of $f_{cat}(t)$ and
the scaling laws that are summarized in Fig. \ref{4}. Note how in
panel (A) the exact solution coincides with the asymptotic approximations
as expected.}
\end{figure}

\textbf{Conclusions.}\textcolor{red}{{} }In this letter we took advantage
of the Michaelis-Menten reaction scheme to provide a unified analysis
of the optimal restart problem. The incorporation of restart into
an existing first passage time problem modifies its behavior. The
mean first passage time then becomes a function of the restart rate
and the question of optimality naturally arises. Here, we have developed
a formalism which can be used in order to study the effect of restart
on generic first passage time problems. A prime corollary of our study
is the identification of two regimes at which the optimal restart
rate displays universal behavior in the sense that it is solely governed
by a handful of key parameters (see Fig. \ref{Fig4} for illustration).
The results we have obtained are applicable to many fields. In particular,
we note that randomized computer algorithms \cite{Randomized algorithm-1,Randomized algorithm-2}
often exhibit heavy tailed run time distributions \cite{Heavy-Tailed run times-1,Heavy-Tailed run times-2}
and restart could hence drastically improve performance in these cases
and others \cite{Restart in CS-1,Restart in CS-2}.

The optimal restart problem is intimately related with the idea of
kinetic proofreading. Independently proposed by Hopfield \cite{kinetic proofreading}
and Ninio \cite{Ninio}, and studied by multiple authors since \cite{kpr1,kpr2,kpr3,kpr4,kpr5,kpr6,kpr7,kpr8,kpr9},
the kinetic proofreading scheme suggests a way in which an enzyme
can amplify \emph{small differences} in the unbinding rates of two
substrates---one right, the other one wrong, such that $k_{off}^{wrong}>k_{off}^{right}$---in
order to discriminate them with \emph{high fidelity}. The basic idea
is that by molding the catalysis time distribution,\textcolor{red}{{}
}for example into something that resembles a sharp delay, the ratio
between the right and wrong turnover rates can be made arbitrarily
large. Consider now a case in which one is provided with a desired
profile of the turnover rate as a function of the unbinding (restart)
rate, and is then asked to conjure a catalysis (i.e., first passage)
time distribution that would yield this profile. For example, think
of a scenario in which an enzyme wishes to select only the substrates
whose unbinding rates fall within a small range (band pass filter),
or above/below some cutoff (high/low pass filter). How could this
be done? Quite surprisingly, a formal solution to this highly non-trivial
problem can be readily obtained by solving Eq. (\ref{2}) for $\hat{f}_{cat}(k_{off})$
given $k_{turn}(k_{off})$ \cite{Caution}. This observation paves
the way towards intelligent design of ``Michaelian Filters''---a
concept which we will further develop elsewhere.

\textbf{Acknowledgments.} We gratefully acknowledge support of the
German-Israeli Project Cooperation Program (DIP). Shlomi Reuveni gratefully
acknowledges support from the James S. McDonnell Foundation via its
postdoctoral fellowship in studying complex systems. We thank Haim
Diamant for proposing possible continuations to our work. We thank
Cristopher Moore for turning our attention to applications in computer
science and for referring us to relevant literature. We thank an anonymous
referee for asking whether the optimal restart problem could be solved
for first passage times that are taken from the one sided Lévy distribution.

\end{document}


\noindent \begin{center}
\textbf{\huge{}\uline{Supplementary Material}}\textbf{\LARGE{}}\\
\textbf{\LARGE{} }
\par\end{center}{\LARGE \par}

\noindent \begin{center}
\textbf{\large{}The Michaelis-Menten reaction scheme}\\
\textbf{\large{}~}\\
\textbf{\large{}as a unified approach towards the optimal restart
problem}\\
\textbf{\large{}~~}
\par\end{center}{\large \par}

\noindent \begin{center}
Tal \textcolor{black}{Rot}bart$^{\dagger,*}$, Shlomi Reuveni$^{\ddagger,*}$,
and Michael Urbakh$^{\dagger}$
\par\end{center}

\noindent \begin{center}
\textit{$^{\dagger}$School of Chemistry, Tel-Aviv University, Tel-Aviv
69978, Israel}
\par\end{center}

\noindent \begin{center}
\textit{$^{\ddagger}$Department of Systems Biology, Harvard University,
200 Longwood Avenue, Boston, Massachusetts 02115, USA.}
\par\end{center}

\noindent \begin{center}
$^{*}$\textit{T. Rotbart and S. Reuveni had equal contribution to
this work.}
\par\end{center}

\noindent \begin{center}
~\\
~\\
 Corresponding Author: Shlomi Reuveni
\par\end{center}

\noindent \begin{center}
Email: shlomireuveni@hotmail.com
\par\end{center}

\clearpage{}

\section{\label{sec:Supplementary-Information}Supplementary information for
figures}

\subsection{\label{Details-Fig2}Details of distributions used in Fig. 2}

Turnover rates were calculated for catalysis time distributions with
$\left\langle T_{cat}\right\rangle =1$, and the mean binding time
was taken to be $\left\langle T_{on}\right\rangle =0.1$ in all cases.
The following distributions were used (numbers match lines No. 1-6
in Figure 2):
\begin{enumerate}
\item Log-Normal distribution: 
\[
f_{cat}(t)=\frac{1}{\sqrt{2\pi}\alpha t}exp\left[\frac{-\left(ln(t)-\beta\right)^{2}}{2\alpha^{2}}\right]\,,
\]
with $\alpha=\sqrt{ln(3)}$, and $\beta=-ln(3)/2$. The variance of
this distribution is $\sigma^{2}(T_{cat})=2\,.$ 
\item Double-Exponential distribution (a.k.a. Hyper-Exponential):
\[
f_{cat}(t)=pk_{1}e^{-k_{1}t}+(1-p)k_{2}e^{-k_{2}t}\,,
\]
with $k_{1}=0.5$, $k_{2}=2$, and $p=1/3$. The variance of this
distribution is $\sigma^{2}(T_{cat})=2\,.$ 
\item Weibull distribution: 
\[
f_{cat}(t)=\alpha\beta\left(\beta t\right)^{\alpha-1}e^{-\left(\beta t\right)^{\alpha}}\,,
\]
with $\alpha=0.7209$, and $\beta=1.232$. The variance of this distribution
is $\sigma^{2}(T_{cat})=2\,.$ 
\item Double-Erlang distribution (a.k.a. Hyper-Erlang):
\[
f_{cat}(t)=pk_{1}^{2}te^{-k_{1}t}+(1-p)k_{2}^{2}te^{-k_{2}t}\,,
\]
with $k_{1}=22.8781,$ $k_{2}=1.4184$, and $p=0.31$. The variance
of this distribution is $\sigma^{2}(T_{cat})\simeq1.0612\,.$
\item Double-Exponential distribution (a.k.a. Hyper-Exponential):
\[
f_{cat}(t)=pk_{1}e^{-k_{1}t}+(1-p)k_{2}e^{-k_{2}t}
\]
with $k_{1}=0.5$, $k_{2}=10/9$, and $p=1/11$. The variance of this
distribution is $\sigma^{2}(T_{cat})=1.2$~. 
\item Exponential distribution:
\[
f_{cat}(t)=ke^{-kt}\,,
\]
with $k=1.$ The variance of this distribution is $\sigma^{2}(T_{cat})=1$~.
\end{enumerate}

\subsection{\label{Details-Fig4}Details of distributions used in Fig. 4}

Maximal turnover rates were calculated for the following catalysis
time distributions (by panel order):
\begin{description}
\item [{A.}] Double-Exponential distribution:
\[
f_{cat}(t)=pk_{1}e^{-k_{1}t}+(1-p)k_{2}e^{-k_{2}t}\,,
\]
with $k_{1}=1$, $k_{2}=10.25$ and $p=0.1025$. 
\item [{B.}] Triple-Exponential distribution:
\[
f_{cat}(t)=p_{1}k_{1}e^{-k_{1}t}+p_{2}k_{2}e^{-k_{2}t}+\left(1-p_{1}-p_{2}\right)k_{3}e^{-k_{3}t}\,,
\]
with $k_{1}=1,$ $k_{2}=6,$ $k_{3}=15,$ $p_{1}=0.1$ and $p_{2}=0.3$.
\item [{C.}] Gamma distribution:
\[
f_{cat}(t)=\frac{\beta}{\Gamma(\alpha)}\left(\beta t\right)^{\alpha-1}e^{-\beta t}\,,
\]
with $\alpha=0.1942$ and $\beta=1.0221$. 
\item [{D.}] Levi-Smirnoff distribution:
\[
f_{cat}(t)=\sqrt{\frac{\tau}{4D\pi t^{3}}}exp\left(\frac{-\tau}{4Dt}\right)\,,
\]
with $D=1$ and $\tau=0.5$. 
\end{description}
\newpage{}

\section{Supplementary Fig. S1}

\begin{figure}[H]
\noindent \begin{centering}
~\\
\includegraphics[scale=0.4]{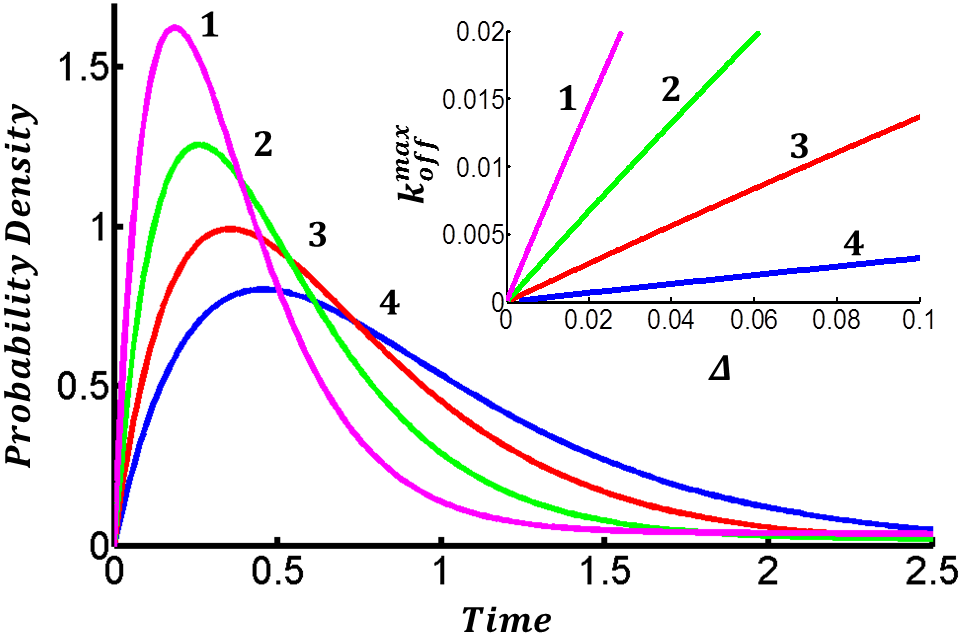}
\par\end{centering}

\noindent \centering{}~\\
\begin{minipage}[t]{1\columnwidth}%
\textbf{Asymptotics of the optimal restart problem, at $\left\langle T_{on}\right\rangle \approx\Psi_{0}$,
is governed by the first three moments of the catalysis time distribution.
}Distributions No. 1-4 are all of type Double Erlang with $\left\langle T_{cat}\right\rangle =1$
and $\sigma(T_{cat})=\sqrt{3}$. They differ, however, by having increasing
skewnesses---$R_{0}=\{0.7,\,1.5,\,3.5,\,14.8\}$ respectively (see
below). \textbf{Inset.} When $\left\langle T_{on}\right\rangle \approx\Psi_{0}$,
$k_{max}$ is linear in $\triangle=\left\langle T_{on}\right\rangle ^{-1}-\Psi_{0}^{-1}$
and inversely proportional to $R_{0}$. As $R_{0}$ increases, the
maximizing unbinding rate becomes lower as to avoid premature termination
of the catalytic process. %
\end{minipage}
\end{figure}

\subsection{Details of distributions used in Fig. S1}

Distributions No. 1-4 are all of type Double-Erlang 
\[
f_{cat}(t)=pk_{1}^{2}te^{-k_{1}t}+(1-p)k_{2}^{2}te^{-k_{2}t}\,.
\]
The following parameters were used:

\noindent \begin{center}
\begin{tabular}{|c|c|c|c|c|c|c|}
\hline 
 & $p$ & $k_{1}$ & $k_{2}$ & $\left\langle T_{cat}\right\rangle $ & $\left\langle T_{cat}^{2}\right\rangle $ & $R_{0}$\tabularnewline
\hline 
\hline 
1 & $1.93*10^{-1}$ & $0.55$ & $5.437$ & $1$ & $4$ & $0.670$\tabularnewline
\hline 
2 & $1.23*10^{-1}$ & $0.45$ & $3.875$ & $1$ & $4$ & $1.467$\tabularnewline
\hline 
3 & $4.94*10^{-2}$ & $0.3$ & $2.833$ & $1$ & $4$ & $3.477$\tabularnewline
\hline 
4 & $4.6*10^{-3}$ & $0.1$ & $2.19$ & $1$ & $4$ & $14.76$\tabularnewline
\hline 
\end{tabular}\newpage{}
\par\end{center}

\section{Supplementary information for text}

\subsection{\label{Derivation-of-Eq3}Derivation of Eq. 3 in the main text}

Before we derive Eq. 3, and the statements which follow it, let us
first obtain a preliminary result. Unbinding is said to be excitatory
whenever $\partial k_{turn}/\partial k_{off}>0$. Utilizing the definition
of $k_{turn}$ (Eq. 2 in the main text), we see that this condition
is equivalent to 
\begin{equation}
\frac{\hat{f}_{cat}(k_{off})\left(1-\hat{f}_{cat}(k_{off})\right)+k_{off}\left(1+k_{off}\left\langle T_{on}\right\rangle \right)\frac{\partial\hat{f}_{cat}(k_{off})}{\partial k_{off}}}{\left(\left\langle T_{on}\right\rangle k_{off}+\left[1-\hat{f}_{cat}(k_{off})\right]\right)^{2}}>0\,.\label{A1}
\end{equation}
Noting that the denominator on the left hand side of Eq. (\ref{A1})
is always positive, we multiply both sides by it and rearrange to
get

\begin{equation}
\hat{f}_{cat}(k_{off})\left(1-\hat{f}_{cat}(k_{off})\right)>-k_{off}\left(1+k_{off}\left\langle T_{on}\right\rangle \right)\frac{\partial\hat{f}_{cat}(k_{off})}{\partial k_{off}}\,.\label{A2}
\end{equation}

The Laplace transform of a probability density function is monotonically
decreasing in $k_{off}$. We can thus take Eq. (\ref{A2}) and divide
both sides by $-\partial\hat{f}_{cat}(k_{off})/\partial k_{off}$
to get 

\begin{equation}
\frac{\hat{f}_{cat}(k_{off})\left(1-\hat{f}_{cat}(k_{off})\right)}{-\frac{\partial\hat{f}_{cat}(k_{off})}{\partial k_{off}}}>k_{off}\left(1+k_{off}\left\langle T_{on}\right\rangle \right)\,.\label{A3}
\end{equation}
Further noting that the unbinding rate is always positive, we rearrange
Eq. (\ref{A3}) to show that the following condition 

\begin{equation}
\frac{\hat{f}_{cat}(k_{off})\left(\hat{f}_{cat}(k_{off})-1\right)}{\frac{\partial\hat{f}_{cat}(k_{off})}{\partial k_{off}}k_{off}^{2}}-\frac{1}{k_{off}}>\left\langle T_{on}\right\rangle \,,\label{A4}
\end{equation}
is equivalent to the existence of excitatory unbinding. 

Examining the left hand side of Eq. (\ref{A4}), we define

\begin{equation}
\Psi(k)\equiv\frac{\hat{f}_{cat}(k)\left(\hat{f}_{cat}(k)-1\right)}{\frac{\partial\hat{f}_{cat}(k)}{\partial k}k^{2}}-\frac{1}{k}\,,\label{eq: Psi}
\end{equation}
and conclude that 

\begin{equation}
\Psi(k_{off})>\left\langle T_{on}\right\rangle \Longleftrightarrow\frac{\partial k_{turn}}{\partial k_{off}}>0\,.\label{A5}
\end{equation}
A similar set of steps can then be applied to show that in the case
of inhibitory unbinding one gets 

\begin{equation}
\Psi(k_{off})<\left\langle T_{on}\right\rangle \Longleftrightarrow\frac{\partial k_{turn}}{\partial k_{off}}<0\,,\label{A6}
\end{equation}
as illustrated in the figure below.\textcolor{red}{{} }

Replacing inequalities with equalities in the above derivation, one
can readily convince himself that 
\begin{equation}
\Psi(k_{off})=\left\langle T_{on}\right\rangle \Longleftrightarrow\frac{\partial k_{turn}}{\partial k_{off}}=0\,,\label{eq:Psi=00003DTon}
\end{equation}
i.e., the stationary points of $k_{turn}$ are the solutions of $\Psi(k_{off})=\left\langle T_{on}\right\rangle $---as
is stated by Eq. 3 in the main text. We furthermore note that a stationary
point of $k_{turn}$ for which $\Psi'(k)<0$ ($\Psi'(k)>0$) is a
a local maximum (minimum) of the turnover rate. Indeed, the amalgamation
of Eqs. (\ref{A5}) and (\ref{A6}) asserts that $\partial k_{turn}/\partial k_{off}$
is positive (negative) if and only if $\Psi(k_{off})-\left\langle T_{on}\right\rangle $
is positive (negative). Since at a stationary point $\Psi(k)-\left\langle T_{on}\right\rangle =0$,
$\Psi'(k)<0$ ($\Psi'(k)>0$) implies that at this point $\partial k_{turn}/\partial k_{off}$
transitions from being positive (negative) to being negative (positive),
thus resulting in a local maximum (local minimum). 
\begin{figure}
\noindent \centering{}\includegraphics[scale=0.6]{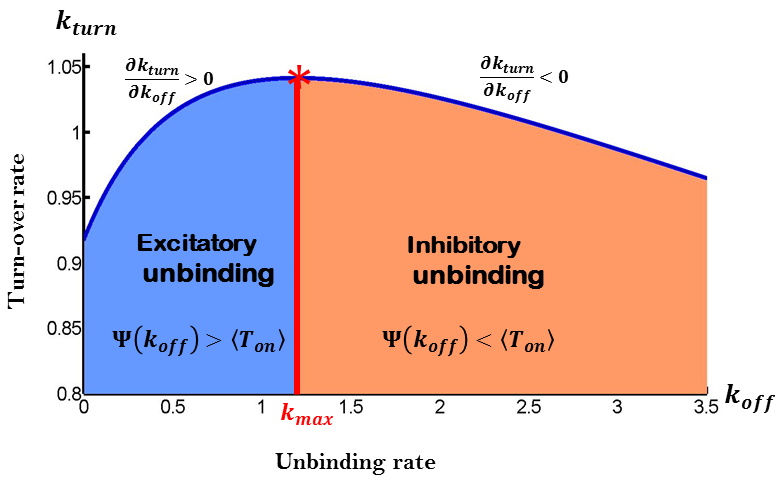}
\end{figure}

\newpage{}

\subsection{Approximating $\Psi(k)$ }

\subsubsection{A general recipe }

Before obtaining approximations for $\Psi(k)$ at particular limits,
we give a recipe illustrating how this could be done in general. Consider
the limit $k\rightarrow k_{lim}$. An approximation for $\Psi(k)$,
at this limit, can be found by taking the following steps: 
\begin{enumerate}
\item Find the third order power expansion of the Laplace transform $\hat{f}_{cat}(k)$
in the limit $k\rightarrow k_{lim}$ (going to third order is required
in order to capture the first order behavior of $\Psi(k)$). If the
Laplace transform is not analytic move to step number (5) below. 
\item Replace the third order power expansion of $\hat{f}_{cat}(k)$ with
the following rational function approximation 
\begin{equation}
\hat{f}_{cat}(k)\simeq\frac{1+ak}{1+bk+ck^{2}}\,.\label{eq:1.9}
\end{equation}
Choose the parameters $a$, $b$, and $c$ such that the third order
power expansion of the right hand side of Eq. (\ref{eq:1.9}) agrees
with that found in step (1) above. 
\item Substitute $\hat{f}_{cat}(k)$ in the definition of $\Psi(k)$ (see
Eq. \ref{eq: Psi} above) with its approximation as given by Eq. (\ref{eq:1.9}). 
\item Note that, under the approximation made in step (3), $\Psi(k)=\left\langle T_{on}\right\rangle $
(Eq. 3 in the main text) is a quadratic equation. Solve it.
\item In cases where a power expansion of $\hat{f}_{cat}(k)$ does not exist,
leading order behavior must be extracted in a different manner. In
some cases, approximate solutions to Eq. 3 in the main text can still
be found. Examples are given below.
\end{enumerate}

\subsubsection{Approximating $\Psi(k)$ in the $k\rightarrow0$ limit }

\textbf{1. Power expansion.} When all moments of the catalysis time
distribution are finite $\hat{f}_{cat}(k)$ can be written in the
following form \cite{First Steps in Random Walks}

\begin{equation}
\hat{f}_{cat}(k)=\sum_{n=0}^{\infty}\frac{M_{n}}{n!}\left(-k\right)^{n}\ ,\label{eq:1.10}
\end{equation}
where $M_{n}=\int_{0}^{\infty}f_{cat}(t)t^{n}dt$ is the $n^{th}$
moment of the distribution. Truncating the infinite series in Eq.
(\ref{eq:1.10}) to obtain a third order power expansion we have

\begin{equation}
\hat{f}_{cat}(k)\simeq1-M_{1}k+\frac{M_{2}}{2}k^{2}-\frac{M_{3}}{6}k^{3}\,.\label{eq:1.11}
\end{equation}

\noindent \begin{flushleft}
\textbf{2. Rational function approximation. }Note that to third order\textbf{
\begin{equation}
\begin{array}{l}
\frac{1+ak}{1+bk+ck^{2}}\simeq\\
\text{ }\\
1-(b-a)k+(b^{2}-ab-c)k^{2}-(b^{3}-ab^{2}+ac-2bc)k^{3}\,.
\end{array}\label{eq:1.12}
\end{equation}
}Equating corresponding coefficients for the power expansions that
appear on the right hand side of Eqs. (\ref{eq:1.11}) and (\ref{eq:1.12}),
one can write equations for the set $\{a,b,c\}$ and solve them to
obtain 
\begin{equation}
a=\frac{6M_{1}(M_{2}-M_{1}^{2})-M_{3}}{3(2M_{1}^{2}-M_{2})},\, b=\frac{3M_{1}M_{2}-M_{3}}{3(2M_{1}^{2}-M_{2})},\, c=\frac{3M_{2}^{2}-3M_{1}M_{3}}{6(2M_{1}^{2}-M_{2})}\,.
\end{equation}
Substituting these parameters into the right hand side of Eq. (\ref{eq:1.9})
we obtain the following rational function approximation
\par\end{flushleft}

\begin{equation}
\hat{f}_{cat}(k)\simeq\frac{1+\left(\frac{6M_{1}(M_{2}-M_{1}^{2})-M_{3}}{3(2M_{1}^{2}-M_{2})}\right)k}{1+\left(\frac{3M_{1}M_{2}-M_{3}}{3(2M_{1}^{2}-M_{2})}\right)k+\left(\frac{3M_{2}^{2}-3M_{1}M_{3}}{6(2M_{1}^{2}-M_{2})}\right)k^{2}}\,.\label{eq:Laplace small k}
\end{equation}

\noindent \begin{flushleft}
\textbf{3. Approximating $\Psi(k)$. }Substituting Eq. (\ref{eq:Laplace small k})
into Eq. (\ref{eq: Psi}) we get\textbf{
\begin{equation}
\begin{array}{l}
\frac{1}{\Psi(k)}\simeq\left[\frac{2M_{1}}{M_{2}-2M_{1}^{2}}\right]+2\left[\frac{2M_{1}M_{3}-3M_{2}^{2}}{3\left(M_{2}-2M_{1}^{2}\right)^{2}}\right]k\\
\text{ }\\
+\left[\frac{\left(6M_{1}^{3}-6M_{2}M_{1}+M_{3}\right)\left(2M_{1}M_{3}-3M_{2}^{2}\right)}{9\left(M_{2}-2M_{1}^{2}\right){}^{3}}\right]k^{2}\,.
\end{array}\label{eq:Psi small explicite}
\end{equation}
} Defining 
\par\end{flushleft}

\begin{equation}
\Psi_{0}\equiv\frac{M_{2}-2M_{1}^{2}}{2M_{1}}\,,
\end{equation}
and
\begin{equation}
R_{0}\equiv\frac{2M_{1}M_{3}-3M_{2}^{2}}{3\left(M_{2}-2M_{1}^{2}\right)^{2}}\,,
\end{equation}
we rewrite Eq. (\ref{eq:Psi small explicite}) in a more compact form 

\begin{equation}
\Psi(k)\simeq\Psi_{0}\left(1+2R_{0}\left(\Psi_{0}k\right)+\left[R_{0}(1+R_{0})\right]\left(\Psi_{0}k\right)^{2}\right)^{-1}\,.\label{eq:1.18}
\end{equation}

Examining Eq. (\ref{eq:1.18}), it can be seen that its right hand
side depends on two parameters only $\Psi_{0}$ and $R_{0}$. We further
note that $\Psi(k=0)=\Psi_{0}$, and that $\Psi_{0}$ defines a time
scale by which we can normalize other times in our problem. Defining 

\begin{equation}
t=\Psi_{0}t',\, k=\Psi_{0}^{-1}\kappa,\,\Psi(k)=\Psi_{0}\Psi_{norm}(\kappa)\,,\label{eq:1.19}
\end{equation}
we can write a normalized version of Eq. (\ref{eq:1.18}) 

\noindent 
\begin{equation}
\Psi_{norm}(\kappa)\simeq\left(1+2R_{0}\kappa+\left[R_{0}(1+R_{0})\right]\kappa^{2}\right)^{-1}\,,\label{eq:1.20}
\end{equation}
in which $\Psi_{0}$ was eliminated, and $R_{0}$---the only remaining
statistic---is dimensionless. 

\noindent \begin{flushleft}
\textbf{4. Solving the governing equation. }Substituting $\Psi(k)$
in Eq. 3 (main text) with its approximation, as given by Eq. (\ref{eq:1.18}),
we have 
\par\end{flushleft}

\begin{equation}
1+2R_{0}\left(\Psi_{0}k\right)+\left[R_{0}(1+R_{0})\right]\left(\Psi_{0}k\right)^{2}=\Psi_{0}/\left\langle T_{on}\right\rangle \,.\label{eq:1.21}
\end{equation}
Equation (\ref{eq:1.21}) is based on a small $k$ approximation of
$\Psi(k)$ and we will hence try to solve in this regime. Small $k$
solutions to Eq. (\ref{eq:1.21}) can only be found in the $\left\langle T_{on}\right\rangle \rightarrow\Psi_{0}$
limit and we therefore see that, since $\left\langle T_{on}\right\rangle \geq0$
by definition, a requirement for their existence is $\Psi_{0}\geq0$
. Two different scenarios should then be distinguished based on the
first order (linear) behavior of the left hand side of Eq. (\ref{eq:1.21}).
If $R_{0}>0$, a small $k$ solution exists only for $0<\left\langle T_{on}\right\rangle <\Psi_{0}$.
On the other hand, if $R_{0}<0$, a small $k$ solution exists only
for $\left\langle T_{on}\right\rangle >\Psi_{0}$. The two solutions
of Eq. (\ref{eq:1.21}) are 

\begin{equation}
k_{ext}^{\pm}=\frac{1}{\left(1+R_{0}\right)\Psi_{0}}\left[\pm\sqrt{\left(1+\frac{1}{R_{0}}\right)\frac{\Psi_{0}}{\left\langle T_{on}\right\rangle }-\frac{1}{R_{0}}}-1\right]\,,\label{eq:1.22}
\end{equation}
but $k_{ext}^{+}\rightarrow0$ in the limit $\left\langle T_{on}\right\rangle \rightarrow\Psi_{0}$,
while $k_{ext}^{-}$ does not. We therefore choose $k_{ext}^{+}$
and continue our analysis. 
\begin{figure}
\includegraphics[scale=0.46]{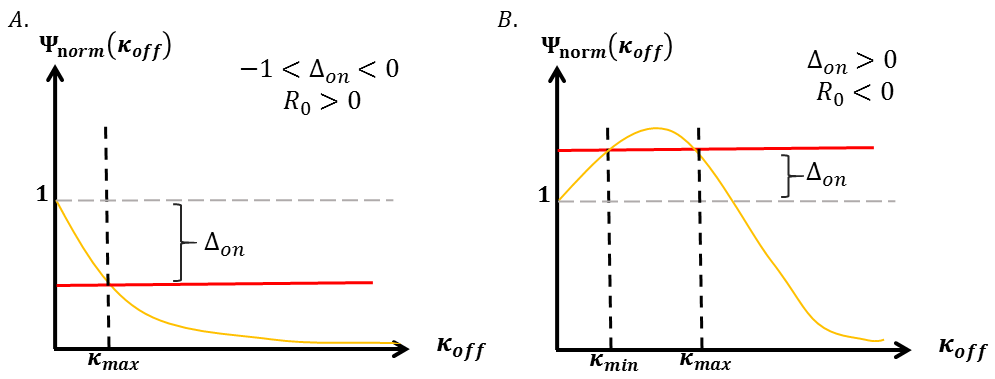}
\end{figure}

Utilizing Eqs. (\ref{eq:1.19}) and (\ref{eq:1.22}) we write $k_{ext}^{+}$
in a normalized form 

\noindent 
\begin{equation}
\kappa_{ext}^{+}=\frac{1}{(1+R_{0})}\left[\sqrt{\frac{1}{1+\Delta_{on}}\frac{R_{0}-\Delta_{on}}{R_{0}}}-1\right]\,,
\end{equation}
where we have defined $\Delta_{on}=\left(\left\langle T_{on}\right\rangle -\Psi_{0}\right)/\Psi_{0}.$
Recalling Eq. (\ref{eq:1.20}), noting that $0<\left\langle T_{on}\right\rangle <\Psi_{0}\Longleftrightarrow-1<\Delta_{on}<0$,
and that $\left\langle T_{on}\right\rangle >\Psi_{0}\Longleftrightarrow\Delta_{on}>0$,
we schematically illustrate the difference between the $R_{0}>0$
case and the $R_{0}<0$ case in the figure above. Most importantly,
we note that from our findings in section (\ref{Derivation-of-Eq3})
it follows that when $R_{0}>0$ ($R_{0}<0$), $\kappa_{ext}^{+}$
is a local maximum (minimum) of the turnover rate.\\

\noindent \textbf{5. Special cases. }The analysis given above was
based on a power expansion of $\hat{f}_{cat}(k)$ in which we assumed
that the first three moments of the catalysis time distribution are
finite. In particular, $\Psi_{0}$ was defined based on the first
two moments of the distribution and is hence not well defined when
these diverge. Divergence of the first two moments can occur if the
PDF has a ``heavy tail'' as happens, for example, when 
\begin{equation}
f_{cat}(t)\simeq\frac{C}{t^{1+\alpha}}\,,
\end{equation}
for $t\gg1$ and $0<\alpha<2$. However, even when this is the case,
one can still say something about the behavior of $\Psi(k)$ at small
$k$, and we distinguish between two different cases. 

When $0<\alpha<1$, the first and second moments of the catalysis
time distribution both diverge. A small $k$ approximation of the
Laplace transform is then given by \cite{First Steps in Random Walks}
\begin{equation}
\hat{f}_{cat}(k)\simeq1-\frac{C\cdot\Gamma(1-\alpha)}{\alpha}k^{\alpha}\,,\label{eq:1.25}
\end{equation}
where $\Gamma(x)$ is the Gamma function. Substituting Eq. (\ref{eq:1.25})
into Eq. (\ref{eq: Psi}) we see that to first order in this limit 

\begin{equation}
\Psi(k)\simeq\frac{1-\alpha}{\alpha}\frac{1}{k}\,.
\end{equation}
Neglecting lower order corrections, we see that $\Psi'(k)<0$ and
conclude that as $\left\langle T_{on}\right\rangle \rightarrow\infty$
the following asymptotics holds 
\begin{equation}
k_{off}^{max}\simeq\frac{1-\alpha}{\alpha}\frac{1}{\left\langle T_{on}\right\rangle }\,.
\end{equation}

When $1<\alpha<2$, the first moment of the catalysis time distribution
is finite but the second moment diverges. A small $k$ approximation
of the Laplace transform is then given by \cite{First Steps in Random Walks}

\begin{equation}
\hat{f}_{cat}(k)\simeq1-\left\langle T_{cat}\right\rangle k+\frac{C\cdot\Gamma(2-\alpha)}{\alpha(\alpha-1)}k^{\alpha}\,.\label{eq:1.28}
\end{equation}
Substituting Eq. (\ref{eq:1.28}) into Eq. (\ref{eq: Psi}) we see
that in this limit 

\begin{equation}
\Psi(k)\simeq\frac{C\cdot\Gamma(2-\alpha)}{\alpha\left\langle T_{cat}\right\rangle }\frac{1}{k^{2-\alpha}}\,,
\end{equation}
Neglecting lower order corrections,we see that $\Psi'(k)<0$ and conclude
that as $\left\langle T_{on}\right\rangle \rightarrow\infty$ the
following asymptotics holds 
\begin{equation}
k_{off}^{max}\simeq\left(\frac{C\cdot\Gamma(2-\alpha)}{\alpha\left\langle T_{cat}\right\rangle \left\langle T_{on}\right\rangle }\right)^{\frac{1}{2-\alpha}}\,.
\end{equation}

\subsubsection{Approximating $\Psi(k)$ in the $k\rightarrow\infty$ limit }

\textbf{1. Power expansion. }When the PDF of the catalysis time distribution
is analytic in the vicinity of $t=0$ we can Taylor expand it 
\begin{equation}
f_{cat}(t)=\sum_{n=0}^{\infty}\left[\left.\left(\frac{\partial}{\partial t}\right)^{n}f(t)\right|_{t=0}\right]\frac{t^{n}}{n!}=\sum_{n=0}^{\infty}\omega_{n+1}\frac{(-t)^{n}}{n!(n+1)!}\,,\label{eq:PDF series}
\end{equation}
where we have defined

\begin{equation}
\omega_{n}\equiv n!\left[\left.\left(-\frac{\partial}{\partial t}\right)^{n-1}f(t)\right|_{t=0}\right]\,,
\end{equation}
from reasons that will soon become clear. The Laplace transform of
such a series can then be written as 

\begin{equation}
\hat{f}_{cat}(t)=\sum_{n=1}^{\infty}\frac{-\omega_{n}}{n!}\left(-\frac{1}{k}\right)^{n}\,.\label{eq:1.33}
\end{equation}
Truncating the infinite series in Eq. (\ref{eq:1.33}) to obtain a
third order power expansion we have
\begin{equation}
\hat{f}_{cat}(k)\simeq\frac{\omega_{1}}{k}-\frac{\omega_{2}}{2k^{2}}+\frac{\omega_{3}}{6}\frac{1}{k^{3}}\,.\label{eq:1.11-1}
\end{equation}

\noindent \begin{flushleft}
\textbf{2. Rational function approximation. }Note that to third order
\par\end{flushleft}

\begin{equation}
\frac{1+ak}{1+bk+ck^{2}}\simeq\frac{a}{c}\left(\frac{1}{k}\right)-\frac{ab-c}{c^{2}}\left(\frac{1}{k}\right)^{2}+\frac{ab^{2}-ac-bc}{c^{3}}\left(\frac{1}{k}\right)^{3}\,.\label{eq:1.12-1}
\end{equation}
Equating corresponding coefficients for the power expansions that
appear on the right hand side of Eqs. (\ref{eq:1.11-1}) and (\ref{eq:1.12-1}),
one can write equations for the set $\{a,b,c\}$ and solve them to
obtain 
\begin{equation}
a=\frac{6\omega_{1}\left(\omega_{2}-2\omega_{1}^{2}\right)}{2\omega_{1}\omega_{3}-3\omega_{2}^{2}},\, b=\frac{2\omega_{3}-6\omega_{1}\omega_{2}}{2\omega_{1}\omega_{3}-3\omega_{2}^{2}},\, c=\frac{6\left(\omega_{2}-2\omega_{1}^{2}\right)}{2\omega_{1}\omega_{3}-3\omega_{2}^{2}}\,.
\end{equation}
Substituting these parameters into the right hand side of Eq. (\ref{eq:1.9})
we obtain the following rational function approximation

\begin{equation}
\hat{f}_{cat}(k)\simeq\frac{1+\left(\frac{6\omega_{1}\left(\omega_{2}-2\omega_{1}^{2}\right)}{2\omega_{1}\omega_{3}-3\omega_{2}^{2}}\right)k}{1+\left(\frac{2\omega_{3}-6\omega_{1}\omega_{2}}{2\omega_{1}\omega_{3}-3\omega_{2}^{2}}\right)k+\left(\frac{6\left(\omega_{2}-2\omega_{1}^{2}\right)}{2\omega_{1}\omega_{3}-3\omega_{2}^{2}}\right)k^{2}}\,.\label{eq:Laplace large k}
\end{equation}

\noindent \begin{flushleft}
\textbf{3. Approximating $\Psi(k)$. }Substituting Eq. (\ref{eq:Laplace large k})
into Eq. (\ref{eq: Psi}) we get\textbf{
\begin{equation}
\begin{array}{l}
\frac{1}{\Psi(k)}\simeq\left[\frac{2\omega_{1}}{\omega_{2}-2\omega_{1}^{2}}\right]k^{2}+2\left[\frac{2\omega_{1}\omega_{3}-3\omega_{2}^{2}}{3\left(\omega_{2}-2\omega_{1}^{2}\right){}^{2}}\right]k\\
\text{ }\\
+\left[\frac{\left(6\omega_{1}^{3}-6\omega_{2}\omega_{1}+\omega_{3}\right)\left(2\omega_{1}\omega_{3}-3\omega_{2}^{2}\right)}{9\left(\omega_{2}-2\omega_{1}^{2}\right){}^{3}}\right]\,.
\end{array}\label{eq:Psi large explicite}
\end{equation}
} Defining 
\par\end{flushleft}

\begin{equation}
\chi_{\infty}\equiv\frac{\omega_{2}-2\omega_{1}^{2}}{2\omega_{1}},\,
\end{equation}
and
\begin{equation}
R_{0}\equiv\frac{2\omega_{1}\omega_{3}-3\omega_{2}^{2}}{3\left(\omega_{2}-2\omega_{1}^{2}\right){}^{2}}\,,
\end{equation}
we rewrite Eq. (\ref{eq:Psi large explicite}) in a more compact form 

\begin{equation}
\Psi(k)\simeq\chi_{\infty}^{-1}\left(\left(\chi_{\infty}^{-1}k\right)^{2}+2R_{\infty}\left(\chi_{\infty}^{-1}k\right)+R_{\infty}\left(R_{\infty}+1\right)\right)^{-1}\,.\label{eq:1.41}
\end{equation}

Examining Eq. (\ref{eq:1.41}), it can be seen that its right hand
side depends on two parameters only $\chi_{\infty}$ and $R_{\infty}$.
We further note that in the limit $k\rightarrow\infty$, $\Psi(k)\simeq\chi_{\infty}k^{-2}$,
and that $\chi_{\infty}^{-1}$ defines a time scale by which we can
normalize other times in our problem. Defining 

\begin{equation}
t=\chi_{0}^{-1}t',\, k=\chi_{0}K,\,\Psi(k)=\chi_{0}^{-1}\Psi_{Norm}(K)\label{eq:1.42}
\end{equation}
we can write a normalized version of Eq. (\ref{eq:1.41})

\begin{equation}
\Psi_{Norm}(K)\simeq\left(K^{2}+2R_{\infty}K+R_{\infty}\left(R_{\infty}+1\right)\right)^{-1}\,,\label{eq: Psi Norm}
\end{equation}
in which $\chi_{\infty}$ was eliminated, and $R_{\infty}$---the
only remaining statistic---is dimensionless. 

\noindent \begin{flushleft}
\textbf{4. Solving the governing equation. }Substituting $\Psi(k)$
in Eq. 3 (main text) with its approximation, as given by Eq. (\ref{eq:1.41}),
we have 
\par\end{flushleft}

\begin{equation}
\frac{\chi_{\infty}/k^{2}}{1+2R_{\infty}\chi_{\infty}/k+R_{\infty}\left(R_{\infty}+1\right)(\chi_{\infty}/k)^{2}}=\left\langle T_{on}\right\rangle \,.\label{eq:1.44}
\end{equation}
Equation (\ref{eq:1.44}) is based on a large $k$ approximation of
$\Psi(k)$ and we will hence try to solve it in this regime. Large
$k$ solutions to Eq. (\ref{eq:1.44}) can only be found in the $\left\langle T_{on}\right\rangle \rightarrow0$
limit. Two different scenarios, schematically illustrated in the figure
above, should then be distinguished based on the leading order behavior
of the left hand side of Eq. (\ref{eq:1.44}). 
\begin{figure}
\includegraphics[scale=0.37]{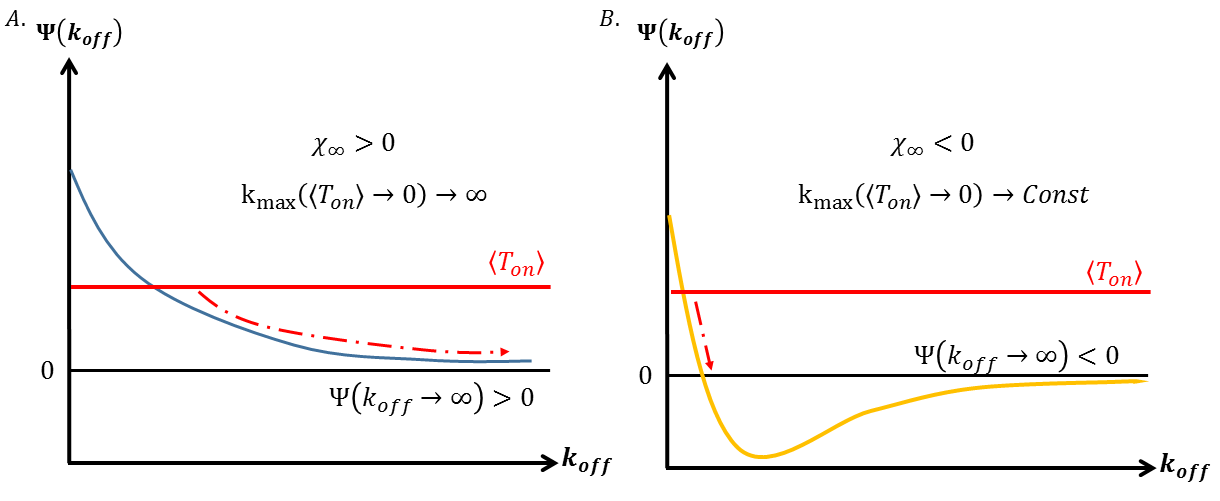}
\end{figure}

When $\chi_{\infty}>0$, $\underset{k\rightarrow\infty}{lim}\Psi(k)=0^{+}$,
i.e., $\Psi(k)>0$ as it approaches zero from above. The two solutions
of Eq. (\ref{eq:1.44}) are then

\begin{equation}
k_{ext}^{\pm}=\chi_{\infty}\left[\pm\sqrt{\frac{1}{\chi_{\infty}\left\langle T_{on}\right\rangle }-R_{\infty}}-R_{\infty}\right]\,,\label{eq:1.45}
\end{equation}
but $k_{ext}^{+}\rightarrow\infty$ in the limit $\left\langle T_{on}\right\rangle \rightarrow0$,
while $k_{ext}^{-}$ does not. We therefore choose $k_{ext}^{+}$
and continue by noting that since $\Psi'(k)<0$ in this limit the
findings in section (\ref{Derivation-of-Eq3}) assert that $k_{ext}^{+}$
is a local maximum of the turnover rate. Utilizing Eqs. (\ref{eq:1.42})
and (\ref{eq:1.45}) $k_{ext}^{+}$ can also be written in normalized
form 

\begin{equation}
K_{ext}^{+}=\sqrt{\frac{1}{\delta_{on}}-R_{\infty}}-R_{\infty}\,,
\end{equation}
where we have defined $\delta_{on}=\chi_{\infty}\left\langle T_{on}\right\rangle .$ 

When $\chi_{\infty}<0$, $\underset{k\rightarrow\infty}{lim}\Psi(k)=0^{-}$,
i.e., $\Psi(k)<0$ as it approaches zero from below. Since $\left\langle T_{on}\right\rangle \geq0$
by definition, there are no large $k$ solutions to Eq. \ref{eq:1.44}
in this case. It is however possible, that $\Psi(k)$ is positive
for smaller values of $k$, in which case, the continuity of $\Psi(k)$
and the fact that it is negative at large $k$ guaranties that there
is a finite value of $k$, $k_{thresh}$, for which $\Psi(k_{thresh})=0$.
This translates (as visualized in the figure above) to $k_{max}\left(\left\langle T_{on}\right\rangle \rightarrow0\right)\rightarrow k_{thresh}\,.$\\

\noindent \textbf{5. Special cases. }The analysis given above was
based on an expansion of $\hat{f}_{cat}(k)$ in powers of $k^{-1}$.
In particular, when defining $\chi_{\infty}$ we have assumed that
$0<f_{cat}(0)=\omega_{1}<\infty$. In contrast, let us now consider
a power law behavior of the PDF 

\noindent 
\begin{equation}
f_{cat}(t)\simeq\omega_{1}t^{\alpha}\,,\label{eq:1.47}
\end{equation}
where $t\ll1$, $\omega_{1}>0$, $\alpha\neq0,$ and $\alpha>-1$
(the PDF cannot be normalized for $\alpha\leq-1$). Noting that the
large $k$ asymptotics of the Laplace transform is then given by

\noindent 
\begin{equation}
\hat{f}_{cat}(k)\simeq\int_{0}^{\infty}\omega_{1}t^{\alpha}e^{-kt}dt=\omega_{1}\Gamma(1+\alpha)k^{-(1+\alpha)}\,,\label{eq:1.48}
\end{equation}
we substitute Eq. (\ref{eq:1.48}) into Eq. (\ref{eq: Psi}) and find 

\begin{equation}
\Psi(k)\simeq\frac{-\alpha}{\alpha+1}\frac{1}{k}\,,\label{eq:1.49}
\end{equation}
to first order in this limit. 

Examining Eq. (\ref{eq:1.49}), we see that two different cases should
be told apart. When $-1<\alpha<0$, $f_{cat}(t)$ diverges in the
limit $t\rightarrow0$, $\Psi(k)>0$, and $\Psi'(k)<0$. Neglecting
lower order corrections, we conclude that as $\left\langle T_{on}\right\rangle \rightarrow0$
the following asymptotics holds 
\begin{equation}
k_{off}^{max}\simeq\frac{-\alpha}{\alpha+1}\frac{1}{\left\langle T_{on}\right\rangle }\,.
\end{equation}

On the other hand, when $\alpha>0$, $f_{cat}(t)$ zeros in the limit
$t\rightarrow0$, and $\Psi(k)<0$. Since $\left\langle T_{on}\right\rangle \geq0$
by definition, there are no large $k$ solutions to Eq. \ref{eq:1.44}
in this case. It is however possible, that $\Psi(k)$ is positive
for smaller values of $k$, in which case, the continuity of $\Psi(k)$
and the fact that it is negative at large $k$ guaranties that there
is a finite value of $k$, $k_{thresh}$, for which $\Psi(k_{thresh})=0$.
In this case, we once again find $k_{off}^{max}\left(\left\langle T_{on}\right\rangle \rightarrow0\right)\rightarrow k_{thresh}\,.$